\documentclass[]{acmart}

\begin{document}

\title{AliCHI: A Large-scale Multi-modal Dataset and Automated Evaluation Tool for Human-like Dialogue Systems}

\author{Zhiling Luo}
\affiliation{%
	\institution{DAMO Academy, Alibaba Group}
	\streetaddress{1e}
	\city{Hangzhou}
	\country{China}}
\email{godot.lzl@alibaba-inc.com}

\author{Qiankun Shi}
\affiliation{%
  \institution{DAMO Academy, Alibaba Group}
  \country{China}
}

\author{Sha Zhao}
\affiliation{%
  \institution{College of Computer Science, Zhejiang University}
  \streetaddress{1e}
  \city{Hangzhou}
  \country{China}}
\email{szhao@zju.edu.com}

\author{Wei Zhou}
\affiliation{%
  \institution{DAMO Academy, Alibaba Group}
\country{China}}

\author{Haiqing Chen}
\affiliation{%
  \institution{DAMO Academy, Alibaba Group}
\country{China}}

\author{Yuankai Ma}
\affiliation{%
  \institution{DAMO Academy, Alibaba Group}
\country{China}}

\author{Haitao Leng}
\affiliation{%
  \institution{DAMO Academy, Alibaba Group}
\country{China}}

\begin{abstract}
\vspace{-0.2 cm}
A well-designed interactive human-like dialogue system is expected to take actions (e.g. smiling) and respond in a pattern similar to humans.
However, due to the limitation of single-modality (only speech) or small volume of currently public datasets, most dialogue systems can only respond in speech and cannot take human-like actions.
In this work, we build a large-scale multi-modal dataset of human-to-human conversation in a face-to-face fashion, with fine-grained annotations.
The raw data in video format contains 635 dialogue sessions, being collected from 200 participants on designed topics and lasting 52 hours in total.
Moreover, we manually annotated the verbal and non-verbal behaviors in each dialogue session on their start/end timestamp. 
Furthermore, we developed a corresponding evaluation tool for human-like dialogue systems to automatically evaluates the accuracy of two basic tasks, turn-taking prediction, and backchannel prediction, on both time and content.
We have opened the data, the tools will be released at the conference.
\end{abstract}


\maketitle

\section{Introduction}

As a kind of popular human-machine interaction system, human-like dialogue systems (HDS) have been widely applied in many scenarios in our daily life. Generally, HDSs are designed for specific applications, for example, an assistant app in a smartphone, virtual customer service assistant, and even a humanoid robot. They pursue the same goal of behaving like a human as much as possible. To this end, HDSs are required to not only reply users’ request in speech, but also take some non-verbal actions, e.g., smiling, which makes HDS more like humans.

HDS tasks include backchannel prediction\cite{jain2021exploring} \cite{murray2022learning} and turn-taking prediction \cite{hara2018prediction} \cite{yang2022gated}, and they require non-verbal behaviors to train human-like models. 
High-quality datasets play a very import role in most existing algorithms for HDS tasks, especially those involving machine learning approaches. 
Previous datasets proposed in this area, such as Noxi\cite{cafaro2017noxi} and JD dataset\cite{yang2022gated}, mostly focus on specific scenarios like knowledge exchange scenarios or commercial systems, where people do not behave in the same way as they do in daily life. 
Also, some of the datasets containing only the utterance and speeches without the non-verbal behaviors, thus cannot support non-verbal behaviors prediction task.

In order to make HDS more human-like and address the above limitation of current public datasets, in this paper, we propose AliCHI, a large-scale multi-modal dataset of human-to-human conversation in a face-to-face fashion, containing fine-grained annotations, and a corresponding tool for evaluation.
Our contributions are two-fold: 

AliCHI, to our best knowledge, is the largest multi-modal dataset on the human-to-human conversation. There are 635 dialogue sessions of 52 hours, collected from 200 participants using our designed topics. High-precision annotations about behaviour time, verbal content, and non-verbal labels are released for backchannel/turn-taking prediction tasks.

Based on our data, we propose a group of evaluating metrics for above two tasks, and develop an automatic CLI tool to test arbitrary HDS systems. The evaluating results computed by the tool can provide a reference for HDS developers.

\begin{figure}[t]
\centering
\includegraphics[width=12cm]{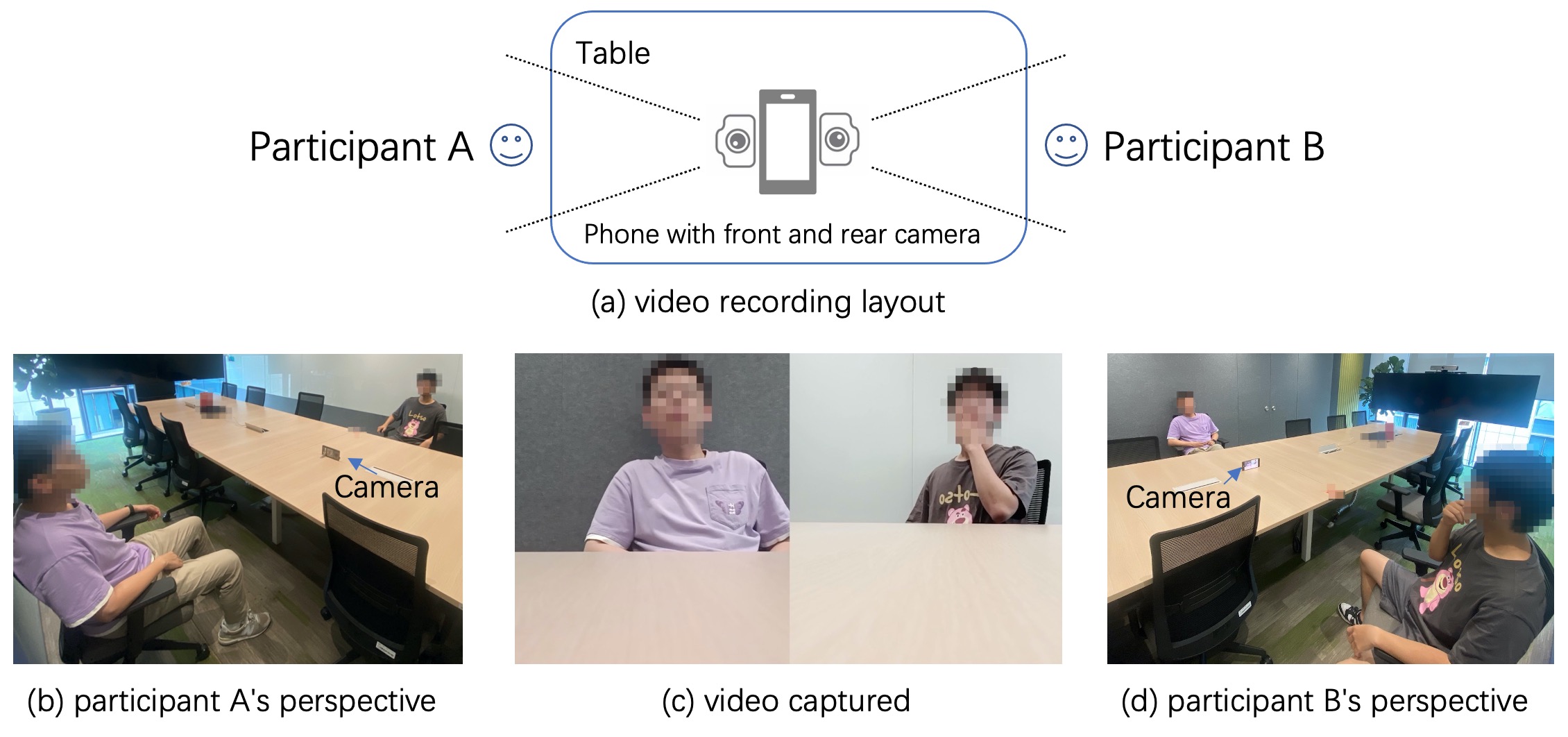}
\caption{Demonstration of the settings in data collection}
\label{fig1}
\end{figure}

\begin{figure}[t]
\centering
\includegraphics[width=12cm]{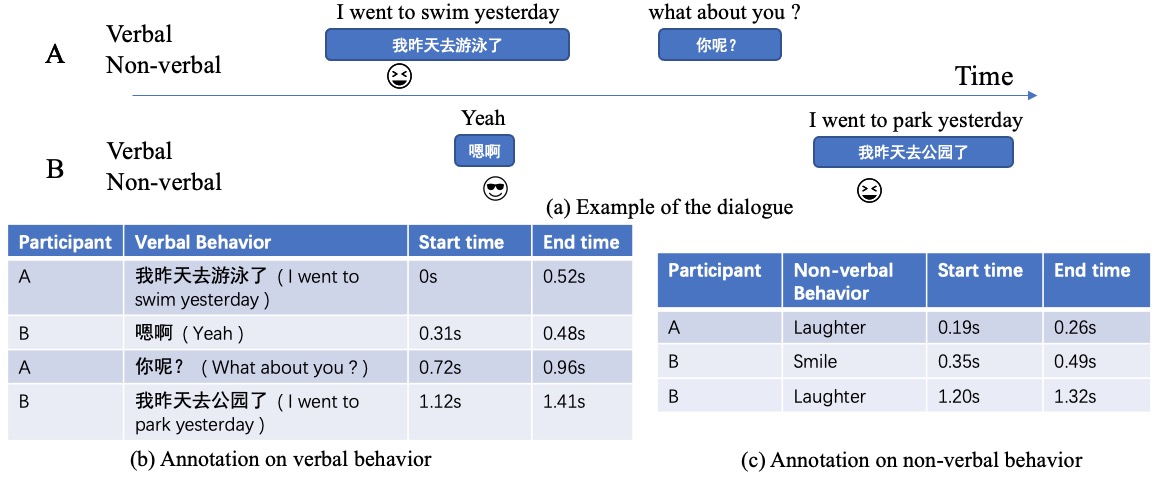}
\caption{Demonstration of the annotations}
\label{fig2}
\end{figure}

\section{Dataset}

\subsection{Design Principle}

To collect the multi-modal face-to-face conversation data, we consider the following principles.

 \textbf{Generic}: The data is expected to reflect the general pattern of both verbal and non-verbal behaviors of humans. 200 participants were invited to collect the data, much more than the number of participants in the NoXi dataset. Such large number of participants can reduce the bias introduced by individual characters to a certain degree. In particular, we design a group of common topics used in conversations, e.g. sports, to further reduce the bias of participants in knowledge. 
We make sure that all the participants have basic knowledge of these topics.

\textbf{Synchronized}: The data contains two video flows from two participants in a dialogue session captured by two cameras. 
These two videos are expected to be synchronized so that the behavior of one participant is easily and exactly aligned to the behavior of another participant.
To reduce the synchronized difference, we leveraged the phone with both front and rear cameras (Xiaomi 10, and HUAWEI Mate40).
Both cameras recorded two videos and generated the fully synchronized video by the phones' inner photograph apps.

\textbf{Natural}: The behaviors between two participants are expected to be natural, rather than rigid with distress.
In each dialogue, the two participants are friends, so that they can talk and act naturally.

\subsection{Data Settings and Collection}

The data is collected in the following settings, shown in Fig. \ref{fig1}a. 
Two participants are required to sit on two sides of a long table ( Fig. \ref{fig1}b, \ref{fig1}d).
The phone with two cameras is placed in the center of the table, ensuring that the two participants are captured in the center of a camera view plane (Fig. \ref{fig1}c).
In this setting, the upper part of the body of the two participants can be captured, thus we can capture many non-verbal behaviors, such as hand gestures.


The data collection process is conducted in sessions.
For each session, two participants choose a topic from our prepared topics (travel, sports, etc.) for the conversation.
The cameras start to capture when they begin to talk about the topic.
The conversation ends if one participant claims that they have finished the talk, and the data collection stops for this session. 
Each session lasts about 5 minutes, and 92\% of sessions last from 3 to 9 minutes. The longest one lasts about 16 minutes, while the shortest one lasts about 2 minutes. By repeating the sessions, we collected 635 sessions in total in the format of video and the time length is 52 hours in total.


\subsection{Data Annotation}



Due to privacy protection policy, we cannot directly release the raw videos which contains the participants' faces. Moreover, the raw data without any annotations is hard to use. Therefore, we tried our best to annotate the dataset.  
The annotated dataset is already released  \footnote{http://www.bruceluo.net/AliCHI.html} and the annotations include:

\textbf{Utterance from speeches} with timestamp, including free-talking, responses, and even backchannel responses;

\textbf{Non-verbal behavior annotations} from a defined annotation set with timestamp. There are 27 annotations in total from five modalities: expressions (e.g., smile), head movements (e.g., nodding), eye movements (e.g., look up), eyebrow movements (e.g., frown), and hand movements (e.g., arms crossed). 


For the entire dataset, 66912 non-verbal behaviors are annotated in total, in which the behaviors of looking left (10,377), looking right (10,246) and smiling (7,299) happen the most frequently.
For the verbal annotation, we labeled the verbal content and its corresponding start/end timestamp. 
In each session, short pause (silence) periods longer than a threshold are used to divide the session into Inter-Pause Units (IPUs).
We noticed within 0.5s before or after the start or end of an IPU, non-verbal behaviors appeared 57\% more than other times, so a more fine-grained division helps to predict non-verbal behaviors more accurately.
The threshold we use is 150ms, which is smaller than the regular 200ms.
There are 159.85 IPUs per session on average and 101,507 IPUs in all.
An example of the annotation is shown in Fig. \ref{fig2}.




\section{Automatic Evaluation Tool}

  

We also provide a tool to automatically evaluate the human-likeness of a HDS on two basic tasks: backchannel prediction and turn-taking prediction.
Considering the example in Fig. \ref{fig2}a, A said about her/his swimming, at the same time B smiled and said 'yeah'.
Both the smiling behavior and short response 'yeah' from B are known as the \textit{backchannel}.
Backchannels are used by the listeners to acknowledge that the current speaker can hold the turn\cite{hara2018prediction}.
After that, A continued and asked B, and then B replied with laughter.
This response is known as \textit{turn-taking} because B took the turn and knew that A would not say anything and just waited for B's response.

\textbf{Backchannel prediction task} takes A's all behaviors to forecast when and how B would reply. 
Specifically speaking, three variables are required to predict: 1) timestamp, measured by IoU; 2) utterance, measured by Bleu and Rouge, with timestamp given; 3) non-verbal behavior annotation, measured by accuracy, with both timestamp and utterance given.

\textbf{Turn-taking prediction task} takes A's all behaviors to forecast when and how B would take the turn.
Similar to backchannel prediction, it predicts 1) timestamp, 2) utterance, and 3) non-verbal behavior.
Note that turn-taking happens on a pause longer than 150ms, but backchannels can happen at any time.

We propose a tool that automatically calculates the above metrics for backchannel and turn-taking prediction.
This tool contains a simple SDK which is easily embedded in HDS to test above tasks by leveraging our dataset as the ground truth.
The code is implemented in Python, which will be released soon, and available via pip.

\bibliographystyle{ACM-Reference-Format}

\bibliography{sample-base}


\begin{thebibliography}{5}


\ifx \showCODEN    \undefined \def \showCODEN     #1{\unskip}     \fi
\ifx \showDOI      \undefined \def \showDOI       #1{#1}\fi
\ifx \showISBNx    \undefined \def \showISBNx     #1{\unskip}     \fi
\ifx \showISBNxiii \undefined \def \showISBNxiii  #1{\unskip}     \fi
\ifx \showISSN     \undefined \def \showISSN      #1{\unskip}     \fi
\ifx \showLCCN     \undefined \def \showLCCN      #1{\unskip}     \fi
\ifx \shownote     \undefined \def \shownote      #1{#1}          \fi
\ifx \showarticletitle \undefined \def \showarticletitle #1{#1}   \fi
\ifx \showURL      \undefined \def \showURL       {\relax}        \fi
\providecommand\bibfield[2]{#2}
\providecommand\bibinfo[2]{#2}
\providecommand\natexlab[1]{#1}
\providecommand\showeprint[2][]{arXiv:#2}

\bibitem[Cafaro~et.al.(2017)]%
        {cafaro2017noxi}
\bibfield{author}{\bibinfo{person}{Angelo Cafaro~et.al.}}
  \bibinfo{year}{2017}\natexlab{}.
\newblock \showarticletitle{The NoXi database: multimodal recordings of
  mediated novice-expert interactions}. In \bibinfo{booktitle}{\emph{IMCI}}.
  \bibinfo{pages}{350--359}.
\newblock


\bibitem[Hara~et.al.(2018)]%
        {hara2018prediction}
\bibfield{author}{\bibinfo{person}{Kohei Hara~et.al.}}
  \bibinfo{year}{2018}\natexlab{}.
\newblock \showarticletitle{Prediction of turn-taking using multitask learning
  with prediction of backchannels and fillers}.
\newblock \bibinfo{journal}{\emph{Listener}}  \bibinfo{volume}{162}
  (\bibinfo{year}{2018}), \bibinfo{pages}{364}.
\newblock


\bibitem[Jain~et.al.(2021)]%
        {jain2021exploring}
\bibfield{author}{\bibinfo{person}{Vidit Jain~et.al.}}
  \bibinfo{year}{2021}\natexlab{}.
\newblock \showarticletitle{Exploring semi-supervised learning for predicting
  listener backchannels}. In \bibinfo{booktitle}{\emph{Proceedings of the 2021
  CHI}}. \bibinfo{pages}{1--12}.
\newblock


\bibitem[Jiudong~et.al.(2022)]%
        {yang2022gated}
\bibfield{author}{\bibinfo{person}{Yang Jiudong~et.al.}}
  \bibinfo{year}{2022}\natexlab{}.
\newblock \showarticletitle{Gated Multimodal Fusion with Contrastive Learning
  for Turn-Taking Prediction in Human-Robot Dialogue}. In
  \bibinfo{booktitle}{\emph{ICASSP 2022}}. IEEE, \bibinfo{pages}{7747--7751}.
\newblock


\bibitem[Murray~et.al.(2022)]%
        {murray2022learning}
\bibfield{author}{\bibinfo{person}{Michael Murray~et.al.}}
  \bibinfo{year}{2022}\natexlab{}.
\newblock \showarticletitle{Learning backchanneling behaviors for a social
  robot via data augmentation from human-human conversations}. In
  \bibinfo{booktitle}{\emph{Conference on Robot Learning}}. PMLR,
  \bibinfo{pages}{513--525}.
\newblock


\end{thebibliography}

\end{document}